\documentclass[prb,twocolumn,showpacs,preprintnumbers,amsmath,amssymb,superscriptaddress]{revtex4}

\usepackage{graphicx}
\usepackage{amssymb}

\begin{document}

\title{Pair breaking by non-magnetic impurities in the non-centrosymmetric \\superconductor CePt$_3$Si}

\author{M. Nicklas}\affiliation{Max Planck Institute for Chemical Physics of
Solids, D-01187 Dresden, Germany}

\author{F. Steglich}\affiliation{Max Planck Institute for Chemical Physics of
Solids, D-01187 Dresden, Germany}

\author{J.~Knolle}
\affiliation {Max Planck Institute for the Physics of Complex Systems, D-01187 Dresden, Germany}

\author {I.~Eremin}
\affiliation {Max Planck Institute for the Physics of Complex Systems, D-01187 Dresden, Germany}
\affiliation {Institute f\"{u}r Mathematische und Theoretische Physik, TU Braunschweig, D-38106
Braunschweig, Germany}

\author{R. Lackner}\affiliation{Institut f\"{u}r Festk\"{o}rperphysik, Technische
Universit\"{a}t Wien, A-1040 Wien, Austria}
\author{E. Bauer}\affiliation{Institut f\"{u}r Festk\"{o}rperphysik, Technische
Universit\"{a}t Wien, A-1040 Wien, Austria}

\begin{abstract}
We have studied the effect of Ge-substitution and pressure on the heavy-fermion super\-conductor
CePt$_3$Si. Ge-substitution on the Si-site acts as negative chemical pressure leading to an increase
of the unit-cell volume, but also introduces chemical disorder. We carried out electrical resistivity
and AC heat-capacity experiments under hydrostatic pressure on CePt$_3$Si$_{1-x}$Ge$_{x}$ ($x=0$,
$0.06$). Our experiments show that the suppression of superconductivity in CePt$_3$Si$_{1-x}$Ge$_{x}$
is mainly caused by the scattering potential, rather than volume expansion, introduced by the Ge
dopands. The antiferromagnetic order is essentially not affected by the chemical disorder.
\end{abstract}


\date{\today}

\maketitle

The physics of unconventional superconductivity in materials without inversion symmetry like the
non-centro\-symmetric heavy-fermion (HF) super\-conductor CePt$_3$Si,\cite{Bauer04}  has recently
become a subject of growing interest. The lack of inversion symmetry - one of the key symmetries for
Cooper pairing - is responsible for a number of novel properties. In particular, a specific property
of the non-centro\-symmetric super\-conductors is that the spin-orbit coupling qualitatively changes
the nature of single-electron states involved in the Cooper pairing by lifting their spin degeneracy
and splitting the conduction bands.\cite{Gorkov01,Frigeri04,Samokin04} For strong spin-orbit
coupling, {\it i.e.}, when the band splitting exceeds the superconducting (SC) energy scales, Cooper
pairing between electrons with opposite momenta occurs only, if they are from the same non-degenerate
band. This scenario is presumably realized in CePt$_3$Si, since, here, band-structure calculations
yield an energy of the spin-orbit coupling, $E_{SO}\approx1000$~K, which is much larger than the SC
critical temperature $T_c\approx0.75$~K.\cite{Samokin04,Bauer04} In the band picture, the SC order
parameter is given by a set of complex wave functions, one for each band, which are coupled due to
interband Cooper-pair scattering. The overall structure of the gap equations resembles those of
multiband super\-conductors, except that the pairing symmetry is more peculiar. While each order
parameter is an odd function of the momentum, the gap symmetry and the positions of the nodes are
determined by one of the even representations of the point group of the crystal. Re-writing this in
the spin representation, one finds the order parameter to become a mixture of singlet and triplet
components, and the latter appears without any spin-triplet term in the underlying pairing
interaction. This is a consequence of the band splitting as well as the difference between the gap
magnitude and the density of states in the different bands.

Polycrystalline CePt$_3$Si orders antiferromagnetically at $T_N=2.2$~K, while superconductivity
appears below $T_c=0.75$~K.\cite{Bauer04} High quality single crystals, in contrast, are SC below
$\sim0.5$ K, while $T_N$ stays unchanged.\cite{Takeuchi07} Previous pressure studies of CePt$_3$Si
revealed a suppression of $T_N$ with increasing pressure as frequently observed in Ce-based HF
metals.\cite{Yasuda04,Nicklas05,Tateiwa05,Aoki07} However, the signature of $T_N$ in electrical
resistivity ($\rho$) and specific heat ($C$) is lost for pressures above $p^*\approx0.6$~GPa,
indicating a sudden suppression of $T_N$. $T_c$ decreases monotonically with application of pressure
and becomes $T_c=0$ at a critical pressure $p_c\approx1.8$~GPa. On the other hand, the crystal
lattice can be expanded, e.g., by isoelectronic substitution of Si by Ge. In
CePt$_3$Si$_{1-x}$Ge$_{x}$ increasing Ge concentration leads to a monotonic increase of the average
values of both the $a$ and the $c$ axis lattice parameter without changing the anisotropy
significantly.\cite{Bauer05} Therefore, in a first approximation, the effect of doping could be
thought of as negative chemical pressure causing a decrease of the hybridization between the $4f$ and
the conduction electrons. Thus, the Kondo interaction should decrease and the magnetic RKKY
interaction increase, in agreement with the observed linear increase of $T_N$ with increasing Ge
concentration.\cite{Bauer05} In contrast, $T_c$ decreases immediately on Ge
substitution.\cite{Bauer05} Thus, CePt$_3$Si appears to be situated right at the position where $T_c$
attains its maximum in the temperature-pressure ($T-p$) phase diagram. We will address the interplay
of antiferromagnetism and superconductivity in CePt$_3$(Si,Ge) and the effect of non-magnetic
impurities on the SC state by $\rho(T)$ and $C(T)$ experiments under hydrostatic pressure.

High quality poly-crystalline material was prepared by high-frequency melting and subsequent
annealing.\cite{Bauer04} Pressures up to 1.85 GPa were generated in a Cu-Be clamp-type cell using
Flourinert FC-75 as the pressure-transmitting medium. High purity Sn served as pressure gauge.
Temperatures down to 50 mK could be reached in a $^3$He/$^4$He dilution cryostat. The resistivity was
measured by a conventional four-probe AC technique using an LR-700 resistance bridge. In addition, AC
specific-heat measurements were conducted using a Au/AuFe thermocouple as thermometer and a
ruthenium-oxide resistance as heater utilizing a digital lock-in amplifier.

\begin{figure}[t]
\centering
\includegraphics[angle=0,width=75mm,clip]{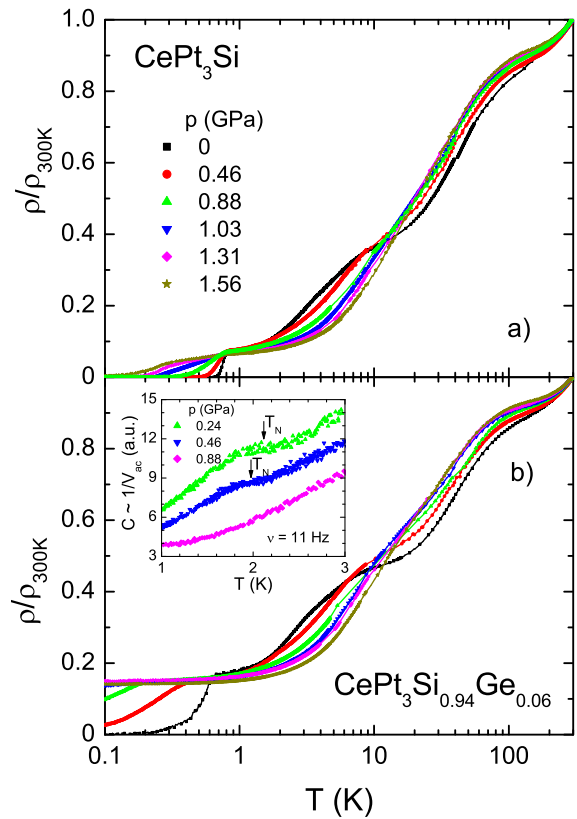}
\caption{\label{rho_high}\label{rho_low}(Color) $\rho(T)$ normalized by $\rho_{\rm 300K}$ of
CePt$_3$Si (a) and CePt$_3$Si$_{0.94}$Ge$_{0.06}$ (b) on a logarithmic temperature scale for
different pressures. Inset of (b): \label{ac-cp}Heat capacity of CePt$_3$Si$_{0.94}$Ge$_{0.06}$
plotted as $C\propto 1/V_{ac}$ vs. $T$ for $p=0.24$, 0.46, and 0.88~GPa. $T_N$, determined by an
equal entropy approximation, is indicated by arrows.}
\end{figure}

Figure \ref{rho_high} shows $\rho(T)$ of  CePt$_3$Si and CePt$_3$Si$_{0.94}$Ge$_{0.06}$ for selected
pressures. At low-$p$ $\rho(T)$ exhibits, for both compounds, the typical features of a Kondo-lattice
system in the presence of a strong crystalline electric-field (CEF) splitting, with shoulders around
10 and 70 K, respectively. The lower shoulder is due to Kondo scattering off the CEF ground-state
doublet, consistent with the Kondo temperature $T_K\approx10$~K, \cite{Bauer04} while scattering off
the fully degenerate $j=5/2$ multiplet of Ce$^{3+}$ leads to the second shoulder at high
temperatures. The position of the high-$T$ shoulder is nearly independent of pressure, which
indicates that the overall CEF splitting is not changing significantly in the pressure range up to
2~GPa. On the other hand, the low-$T$ shoulder shifts strongly to higher temperature with increasing
pressure and finally merges with the high-$T$ shoulder. This behavior is typical for Ce-based
Kondo-lattice compounds where $T_K$ generally increases with pressure. Although, the overall pressure
and temperature dependence of the resistivity is similar for both, pure and Ge-substituted
CePt$_3$Si, the low-$T$ shoulder depends much stronger on pressure in case of
CePt$_3$Si$_{0.94}$Ge$_{0.06}$. At $1.31$~GPa, a clear local shoulder at low temperatures is still
apparent in the Ge-substituted compound, while in the case of pure CePt$_3$Si it has already merged
with the high-$T$ shoulder. Thus the pressure response of $T_K$ in CePt$_3$Si$_{0.94}$Ge$_{0.06}$
exceeds the one in CePt$_3$Si.

At $p=0$, CePt$_3$Si and CePt$_3$Si$_{0.94}$Ge$_{0.06}$ order anti\-ferro\-magnetically at $T_N=2.2$
and 2.8 K, before becoming SC at $T_c=0.75$ and 0.29 K, respectively.\cite{Bauer04,Bauer05} A smooth
change in the curvature indicates the onset of antiferromagnetic (AF) order in $\rho(T)$. The
inflection point of $\rho(T)$ agrees well with $T_N$ obtained from specific-heat data (cf.\ inset of
Fig.~\ref{ac-cp}b). In the following we define $T_N$ from the inflection point in $\rho(T)$. $T_N(p)$
is suppressed with increasing pressure. Above $p^*\approx0.6$ GPa, no signature of the AF transition
is observed anymore in either $\rho(T)$ or $C(T)$ data, suggesting a sudden suppression of $T_N(p)$
in both materials. For CePt$_3$Si similar results have been reported
previously.\cite{Nicklas05,Tateiwa05} Consistent with the stronger pressure dependence of $T_K$ in
the Ge-substituted system, the absolute value of its slope $|{\rm d} T_N(p)/{\rm d} p|_{p=0}$ is
larger than that of the stoichiometric compound [${\rm d} T_{N}(p)/{\rm d} p|_{p=0}{=-1.2\rm~K/GPa}$
and $-0.94{\rm~K/GPa}$, respectively].

\begin{figure}[t]
\centering
\includegraphics[angle=0,width=75mm,clip]{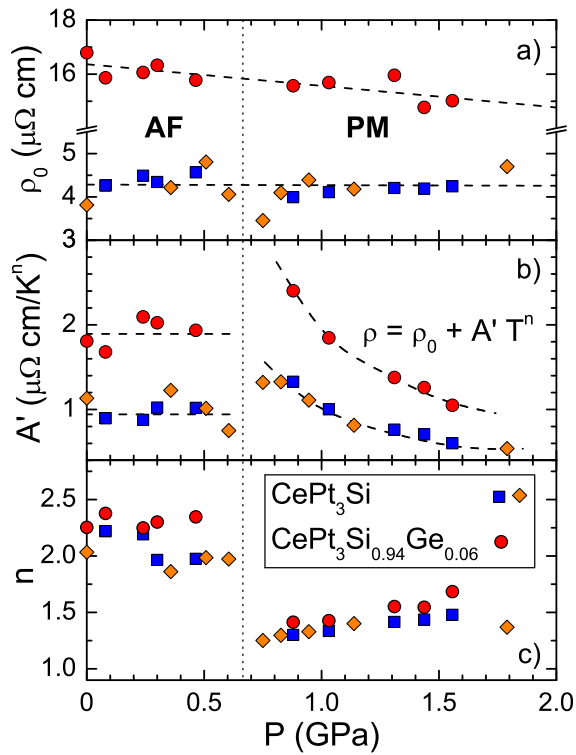}
\caption{\label{parameter}(Color online) Fit parameters, $\rho_0$ (a), $A'$ (b), and $n$ (c) obtained
by fitting the low temperature normal state resistivity using $\rho(T)=\rho_0+A'T^n$. In addition to
the results from this work, for CePt$_3$Si data from Ref.~\onlinecite{Nicklas05} has been analyzed
(diamonds). The vertical dotted line delineates the border between the antiferromagnetic (AF) and the
paramagnetic (PM) regions. Dashed lines are guides to the eye.}
\end{figure}

Figure~\ref{parameter} shows the results of a fit of $\rho(T)=\rho_0+ A' T^n$ to the low-$T$
resistivity data ($T_c\leq T\leq \min\{T_N{\rm , 4\,K}\}$). As expected, the alloying-induced
disorder causes an increase of the residual resistivity, $\rho_0$, by a factor of about 4 from
CePt$_3$Si to CePt$_3$Si$_{0.94}$Ge$_{0.06}$. For both samples $\rho_0(p)$ is nearly pressure
independent; only a weak decrease is observed for the Ge-substituted sample. Especially, there is no
feature in $\rho_0(p)$ at $p^*$ where the AF order disappears. $\rho(T)$ follows a $T^2$ behavior
inside the AF state. Simultaneously with the loss of the signature of the N\'{e}el transition in
resistivity and heat capacity, the temperature dependence of $\rho$ changes drastically for both
compounds in that the pressure dependence of resistivity exponent $n(p)$ exhibits a sharp step from
$n\approx 2$ below $p^*\approx0.6$ GPa to $\approx 1.3$ at higher pressure. Therefore, the heavy
Fermi-liquid (FL) phase ($n=2$) at low pressure becomes unstable against a non-Fermi liquid phase
($n<2$), which exists in the whole pressure range $p^*<p\leq1.88$~GPa, the highest pressure in our
experiment. In the FL phase $A'$ is a measure of the effective quasi-particle -- quasi-particle (QP
-- QP) scattering cross section. $A'$ stays constant in the AF phase and, therefore, the QP -- QP
scattering cross section does not change. In particular, there is no divergence of the QP -- QP
scattering rate or a strong increase of $\rho_0(p)$ as it might be expected on approaching a quantum
critical point.

\begin{figure}[t]
\centering
\includegraphics[angle=0,width=75mm,clip]{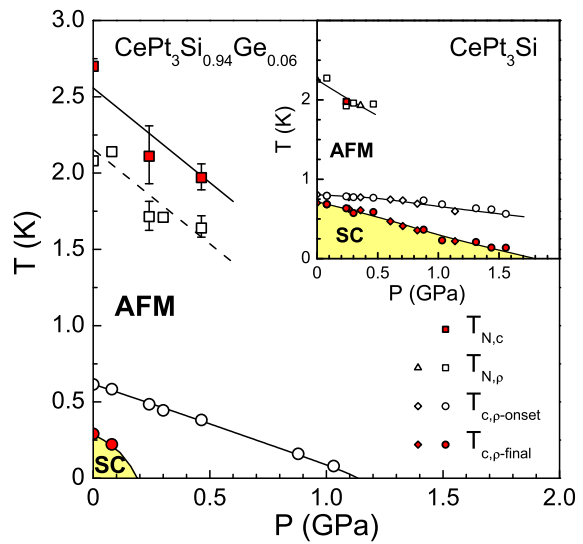}
\caption{\label{pT_PhD}(Color online) $T-p$ phase diagram of CePt$_3$Si$_{0.94}$Ge$_{0.06}$ (main
panel) and CePt$_3$Si (inset). In case of CePt$_3$Si, additional data from
Ref.~\onlinecite{Nicklas05} (triangles and diamonds) have been included.}
\end{figure}

The results of our experiments are summarized in the phase diagram displayed in Fig.~\ref{pT_PhD}.
$T_{c,\rho {\rm\,onset}}$ and $T_{c,\rho {\rm\,final}}$ are taken as the temperatures where the
resistivity is reduced to 90\% and 10\% of its value in the normal state, respectively. In
CePt$_3$Si, $T_{c,\rho {\rm\,final}}$ extrapolates to $T=0$ at about $p\approx1.8$~GPa. Whereas $T_c$
is suppressed upon increasing pressure, the resistive transition successively broadens, i.e. from
$\Delta T=T_{c,\rho {\rm\,onset}}-T_{c,\rho {\rm\,final}}=105$~mK at 0~GPa to $\Delta T=420$~mK at
1.56~GPa. Compared with CePt$_3$Si the SC state in CePt$_3$Si$_{0.94}$Ge$_{0.06}$ responds more
sensitively to external pressure. At $0.24$ GPa, no $\rho=0$ state is observed any more, but the
onset of the SC transition in resistivity is still visible up to $1.03$~GPa. No surprise, for the
alloy, the resistive transition at ambient pressure is found to be already rather broad, $\Delta
T=325$~mK. A recent neutron Lamor-diffraction study on high-quality single crystalline CePt$_3$Si
evidenced a wide distribution of lattice constants of $\approx 10^{-3}$, which might be interpreted
as a wide range of effective pressures across the the sample volume. Consequently, a substantial
width of $\Delta T$ is already observed in CePt$_3$Si.\cite{Ritz09}

Considering the dome-like shape of the SC phase often observed in a $T-p$ phase diagram in HF
super\-conductors, CePt$_3$Si seems to be situated close to the $T_c$-maximum which occurs at a
hypothetical minor negative pressure.\cite{Bauer05} The small initial slope of $T_c(p)$ suggests that
the maximum $T_c$ exceeds $T_c$ at ambient pressure only slightly. It is important to note that
substituting Si by isoelectronic Ge expands the unit-cell volume without changing the electronic
structure significantly.\cite{Bauer05}

Doping with 6\% Ge leads to an increase of the unit-cell volume, $V$, of $\approx0.38\%$ compared
with the stoichiometric compound. Using the bulk modulus $B=162$~GPa,\cite{Ohashi09} this corresponds
to the application of a hypothetical negative pressure of $\Delta p=-0.6$~GPa, resulting in a
reduction of $T_c$ and an increase of $T_N$. Since the volume expansion reduces the 4{\it f} -
conduction electron hybridization and, this way, strengthens the RKKY interaction (while weakening
the Kondo effect), the observed dependence of $T_N$ on the unit-cell volume can be easily explained.
Since the partial Ge-substitution for Si should have no significant effect on the local environment
of the Ce$^{3+}$ ions, it can be expected that disorder has only a minor influence on the magnetic
properties in this material.

The consequence of adding a non-magnetic impurity in a non-centro\-symmetric super\-conductor is far
from being obvious. Theoretical analysis shows that adding non-magnetic impurities
results,\cite{mineev} for weak disorder, in the suppression of $T_c$ for {\it both} conventional as
well as unconventional Cooper pairing. Moreover, for the conventional Cooper pairing, non-magnetic
impurities yield a decrease of $T_c$; superconductivity, {\it however}, will not be destroyed
completely. This is because the origin of the suppression of $T_c$ is interband impurity scattering
which tends to reduce the difference between the gap magnitudes in the two bands. This costs energy
and thus suppresses $T_c$. However, once both gaps have become equal, adding further impurities
should be harmless for superconductivity, which is in striking contrast to our observation that
superconductivity is completely suppressed for a doped sample of $10\%$ Ge.\cite{Bauer05} Thus, the
latter observation points toward an unconventional symmetry of the Cooper pairing in CePt$_3$Si,
involving lines of nodes. The particular symmetry of the order parameter is not yet known in
CePt$_3$Si, however, line nodes occur for any of the $A_2$, $B_1$, or $B_2$ of the $C_{4v}$ group.
Then, for all types of unconventional pairing, the suppression of $T_c$ is described by the universal
Abrikosov-Gor'kov (AG) function,\cite{AG}
$\ln\left[\frac{T_{c}}{T_{c0}}\right] = \Psi \left(\frac12\right) - \Psi \left(\frac12
+\frac{\alpha}{2\pi k_B T_c} \right)$,
where $T_{c0}$ is the SC transition temperature without impurities and $\alpha = \frac{h}{2 \tau}$ is
the pair-breaking parameter. Here, $\tau^{-1}  = 2 \pi n_{imp} N(0) I^2$ is the inverse collision
time resulting for the impurity potential with $n_{imp}$ being the impurity concentration, $N(0)$ is
the density of states at the Fermi level, and $I$ is the scattering potential.

\begin{figure}[!t]
\centering
\includegraphics[angle=0,width=75mm,clip]{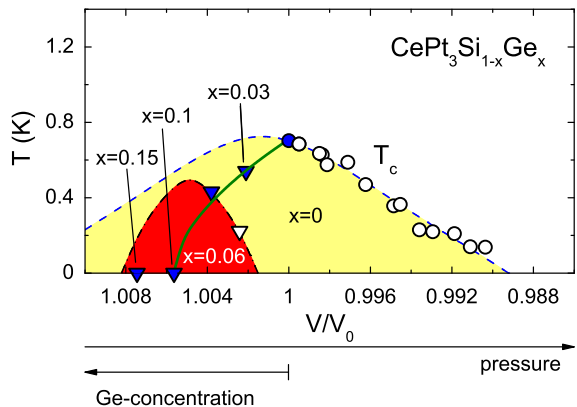}
\caption{\label{VT_PhD}(Color online) Phase diagram of CePt$_3$Si$_{1-x}$Ge$_x$ as function of the
reduced unit-cell volume ($V/V_0)$, with $V_0$ the unit-cell volume of CePt$_3$Si.\cite{volume}
Triangles correspond to $T_c$ of Ge-substituted samples, while circles present $T_c$ of CePt$_3$Si.
Ambient pressure data is indicated by filled symbols, while data under applied pressure by open
symbols. $T_c$ of the Ge-substituted samples at $p=0$ is taken from Ref.~\onlinecite{Bauer05}. The
solid line represents a fit to the data according to the AG function. The dashed and dashed-dotted
lines are suggesting the shape of the SC dome for CePt$_3$Si and CePt$_3$Si$_{0.94}$Ge$_{0.06}$,
respectively, for the case that the SC behavior would be only governed by volume effects. See text
for details.}
\end{figure}

The solid line in Fig.~\ref{VT_PhD} shows the fit of the AG function to our experimental results as a
function of $n_{imp}$ and assuming $N(0) I^2 \sim 1.6 \times 10^{-4}$~eV. The agreement between our
experimental data and the AG theory clearly points toward the unconventional symmetry of the SC order
parameter in CePt$_3$Si which is destroyed by non-magnetic impurities. In addition, we find that the
strength of the potential scattering off the Ge dopands, rather than the Ge-induced expansion of the
average unit-cell volume, more strongly affects $T_c$, quite opposite to the response of $T_N$ to
these parameters.

On applying pressure the SC state in CePt$_3$Si$_{0.94}$Ge$_{0.06}$ is suppressed already at a small
pressure of only $0.2$~GPa. The pressure dependence of $T_c$ is much stronger than in the case of
CePt$_3$Si. Especially, $T_c(p)$ does not exhibit an initial increase as one would expect it in the
simple picture which was first suggested by combining the $p=0$ results on CePt$_3$Si$_{1-x}$Ge$_{x}$
and the pressure studies on CePt$_3$Si in a $T-V$ phase diagram (see also
Fig.~\ref{VT_PhD}).\cite{Bauer05} Our results, however, suggest that the effect of non-magnetic
impurities on the superconductivity cannot be neglected.

In conclusion, our results show that the suppression of $T_c$ on Ge-substitution in CePt$_3$Si is
basically not due to a volume effect, but is caused by scattering processes on non-magnetic
impurities introduced by the Ge-substitution. We have argued that the peculiar effect of non-magnetic
impurities in non-centro\-symmetric super\-conductors plays an important role in destroying
superconductivity in CePt$_3$Si. In addition, we have shown that the SC state in the Ge-substituted
sample is much more sensitive to pressure than in CePt$_3$Si.

Support by COST P16 is acknowledged.


\begin{thebibliography}{99}
\bibitem{volume} The unit-cell volume of CePt$_3$Si$_{x}$Ge$_{1-x}$ 
at $p=0$ was taken from Ref.~\onlinecite{Bauer05}. Pressure was converted to unit-cell volume using
Murnaghan's equation of state with $B=162$~GPa.\cite{Ohashi09}
\end{thebibliography}
\end{document}